\documentclass[12pt]{article}
\usepackage{pic02}
\usepackage{hyperref}
\usepackage{url}
\usepackage{graphicx}
\usepackage{epsfig}
\usepackage{floatflt}

\begin{document}

\newcommand{\snn}{\sqrt{s_{NN}}}
\newcommand{\seff}{\sqrt{s_{\rm eff}}}
\newcommand{\s}{\sqrt{s}}
\newcommand{\pp}{pp}
\newcommand{\pbarp}{\overline{p}p}
\newcommand{\qbarq}{\overline{q}q}
\newcommand{\epem}{e^+e^-}
\newcommand{\nch}{N_{ch}}
\newcommand{\np}{N_{part}}
\newcommand{\nc}{N_{coll}}
\newcommand{\halfnp}{\langle\np/2\rangle}
\newcommand{\ns}{N_{spec}}
\newcommand{\ntot}{\langle\nch\rangle}
\newcommand{\avenp}{\langle\np\rangle}
\newcommand{\as}{\alpha_{s}(s)}
\newcommand{\nubar}{\overline{\nu}}
\newcommand{\ratio}{\ntot/\halfnp}
\newcommand{\etazero}{\eta = 0}
\newcommand{\etaone}{|\eta| < 1}
\newcommand{\dndeta}{d\nch/d\eta}
\newcommand{\dndetazero}{\dndeta|_{\etazero}}
\newcommand{\dndetaone}{\dndeta|_{\etaone}}
\newcommand{\dndetanp}{\dndeta / \halfnp}
\newcommand{\dndetaonp}{\dndeta / \np}
\newcommand{\dndetazeronp}{\dndetazero / \halfnp}
\newcommand{\npp}{n_{pp}}
\newcommand{\vt}{v_{2}}
\newcommand{\pt}{p_T}
\newcommand{\raa}{R_{AA}}
\newcommand{\kstar}{K^\star}
\newcommand{\gevfm}{{\rm GeV}/{\rm fm}^3}

\title{\bf RELATIVISTIC HEAVY ION PHYSICS:\\RESULTS FROM AGS TO RHIC}
\author{
Peter Steinberg\\
\vspace*{.5cm}
{\em Brookhaven National Laboratory, Upton, NY 11973}}
\maketitle

%
%
\begin{figure}[h]
\begin{center}
%
%
%
%
\vspace{4.5cm}
\end{center}
\end{figure}

\baselineskip=14.5pt
\begin{abstract}
High-energy collisions of heavy ions provide a means to study
QCD in a regime of high parton density, and may provide insight
into its phase structure.
Results from the four experiments at RHIC (BRAHMS, PHENIX,
PHOBOS and STAR) are presented, and placed in context with the
lower energy data from the AGS and SPS accelerators.  
The focus is on the insights these measurements provide
into the time history of the collision process.
Taken together, the data point to the creation of
a deconfined state of matter that forms quickly, expands
rapidly and freezes out suddenly.
\end{abstract}
\newpage

\baselineskip=17pt

\section{Introduction}

\begin{figure}[hb]
\begin{minipage}[t]{75mm}
\raisebox{1cm}
{
\includegraphics[width=75mm]{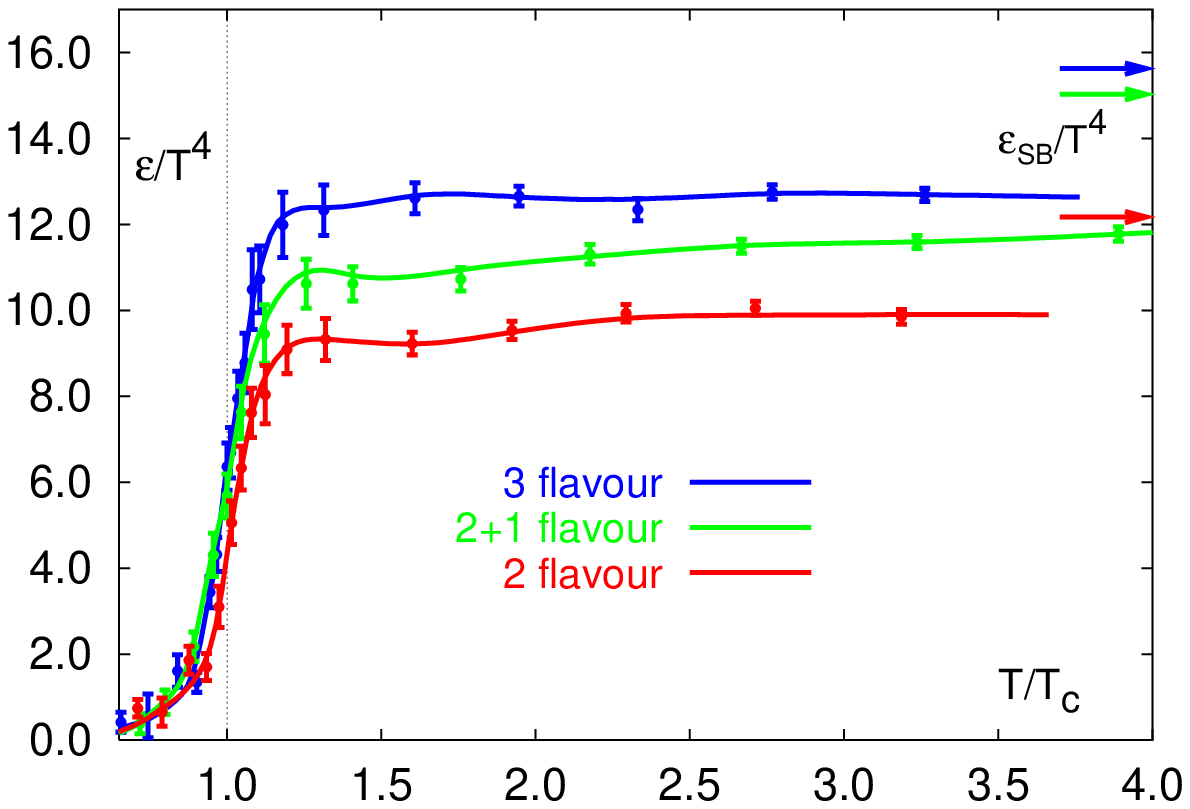}
}
\caption{\it Lattice predictions for the energy 
density as a function of $T/T_c$.}
\label{fig:lattice-pred}
\end{minipage}
\hspace{\fill}
\begin{minipage}[t]{75mm}
\begin{center}
\raisebox{.5cm}
{
\includegraphics[height=60mm]{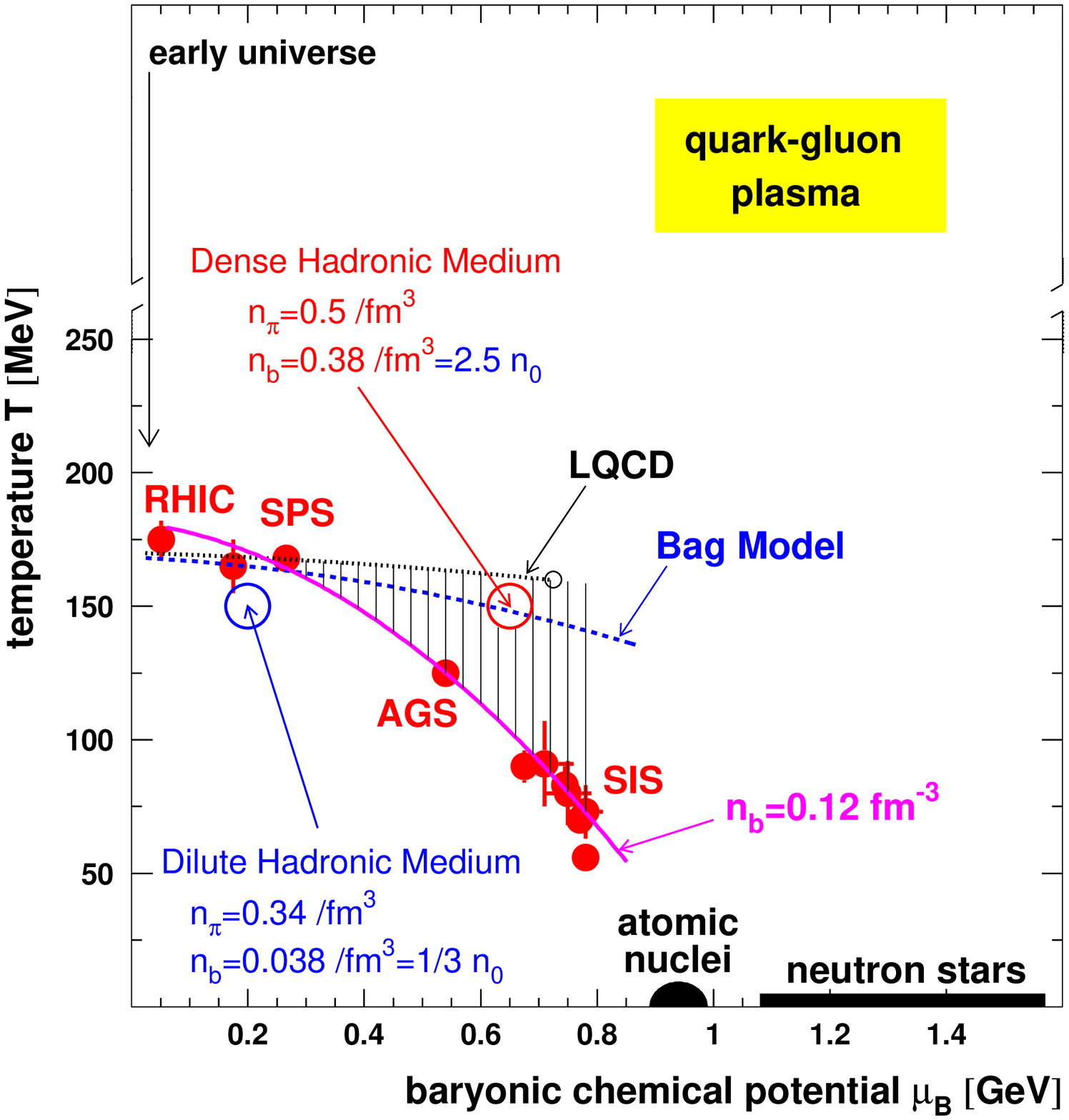}
}
\caption{\it Theoretical expectations for the QCD
phase diagram.}
\label{fig:qgp-phase-diag}
\end{center}
\end{minipage}
\end{figure}

The goal of high energy heavy ion physics is to study QCD in
a regime of high temperature, high density, and
large reaction volumes.  The hope is to
find conclusive evidence that QCD undergoes a phase transition
at a critical temperature from a confined state, where quarks
and gluons are bound in colorless hadron states, to a deconfined
quark-gluon plasma (QGP), 
where quarks and gluons can explore volumes larger than the
typical hadron radius ($R\sim1$ fm).
Lattice calculations, under a variety of assumptions (e.g. number
of quark flavors or $m_s$), make the robust
prediction that the degrees of freedom available to the system
rise rapidly as it is heated through the critical temperature
$T_c \sim 150-200$ MeV \cite{lattice}, as shown in 
Fig. \ref{fig:lattice-pred}.  This corresponds to an energy density
of order $\epsilon_c \sim 1-2$ GeV/fm$^3$.  
There is great interest within the theoretical community whether
these lattice predictions will be confirmed in experiments and,
more generally, whether high temperature QCD can be used to make
quantitative predictions.

While the theoretical questions are compelling in themselves,
the experimental challenges of measuring the properties of
heavy ion collisions in the laboratory imply a related
set of questions that should be addressed before definitive
theoretical conclusions can be made.  The overarching problem
is whether it is possible to discern the various stages in the
time evolution of the system, which evolves from
the initial collisions to the final state hadrons
in times on the order of 10's of fm/c.    To discover a QGP
formed in the early stages of the collision, there must be
observables which maintain information about its properties 
throughout the expansion and cooling of the system and the
final hadronization stage which forms many of the particles that
register in the experimental apparatus.  

While measurables
which address the early times directly (e.g. lepton and photon observables)
are becoming available, the emphasis here will be on 
hadronic observables.
The claim in this proceedings is that the evidence collected
so far points to a picture of the collision dynamics
which is consistent with the formation of a deconfined
state that forms quickly, expands rapidly, and freezes out
suddenly.  There are also interesting comparisons with
charged particle production in $\epem$ collisions that
provides some connections with perturbative QCD phenomenology.

\section{Colliders \& Experiments}


High energy beams of heavy nuclei have been available 
at 
the Brookhaven AGS (Au+Au at the CMS energies per
$NN$ collision from $\snn = 2.5-4.3$ GeV).  
and
the CERN SPS (Pb+Pb collisions from $\snn = 8-17.3$ GeV).
Since then,
a new collider, RHIC, has begun colliding nuclei at higher
energies (up to $\snn = 200$ GeV) with substantial luminosities
(of up to $10^{26} {\rm sec}^{-1}{\rm cm}^{2}$).
The results shown here focus primarily on RHIC data from Au+Au
collisions over a range of energies ($\snn$ = 19.6, 56, 130
and 200 GeV).  Most of the results are from the 130 GeV run, where
the available luminosity limited the $\pt$ reach of many observables.
However, early results from the 200 GeV data are shown here, and some
preliminary results shown at the recent Quark Matter 
conference are also mentioned \cite{QM2002}.

There are four RHIC experiments, which can be 
classified into two groups: the ``large''
experiments, PHENIX and STAR, which are large-volume and large-acceptance
general-purpose detectors, and the ``small'' experiments, BRAHMS
and PHOBOS, which are have more limited acceptance
but cover aspects of the collisions not addressed by the other experiments.  
STAR has a large TPC with coverage
for charged hadrons in $|\eta|<1$ and particle identification 
via $dE/dx$ up to $p_{T}\sim$ 1 GeV.  PHENIX has more limited
hadron coverage ($|\eta|<.35$) but a higher data rate,
extensive particle identification for hadrons, photons, and electrons, 
and forward muon arms.  
PHOBOS has a small-acceptance silicon spectrometer
near $y\sim1$ and $4\pi$ acceptance for multiplicity measurements.
BRAHMS has two movable multi-particle spectrometers,
one at midrapidity, and one covering rapidities out to
$y \sim 4$, offering a
look at identified particle spectra in regions not covered by
any of the other RHIC experiments.  
The four RHIC experiments complement each other in that they
study many different aspects of heavy ion collisions, but
all have hadron coverage near mid-rapidity, which allows detailed
cross-checks between the experiments.

\section{Initial Conditions}

\begin{figure}[h]
\begin{minipage}[t]{65mm}
\includegraphics[width=65mm]{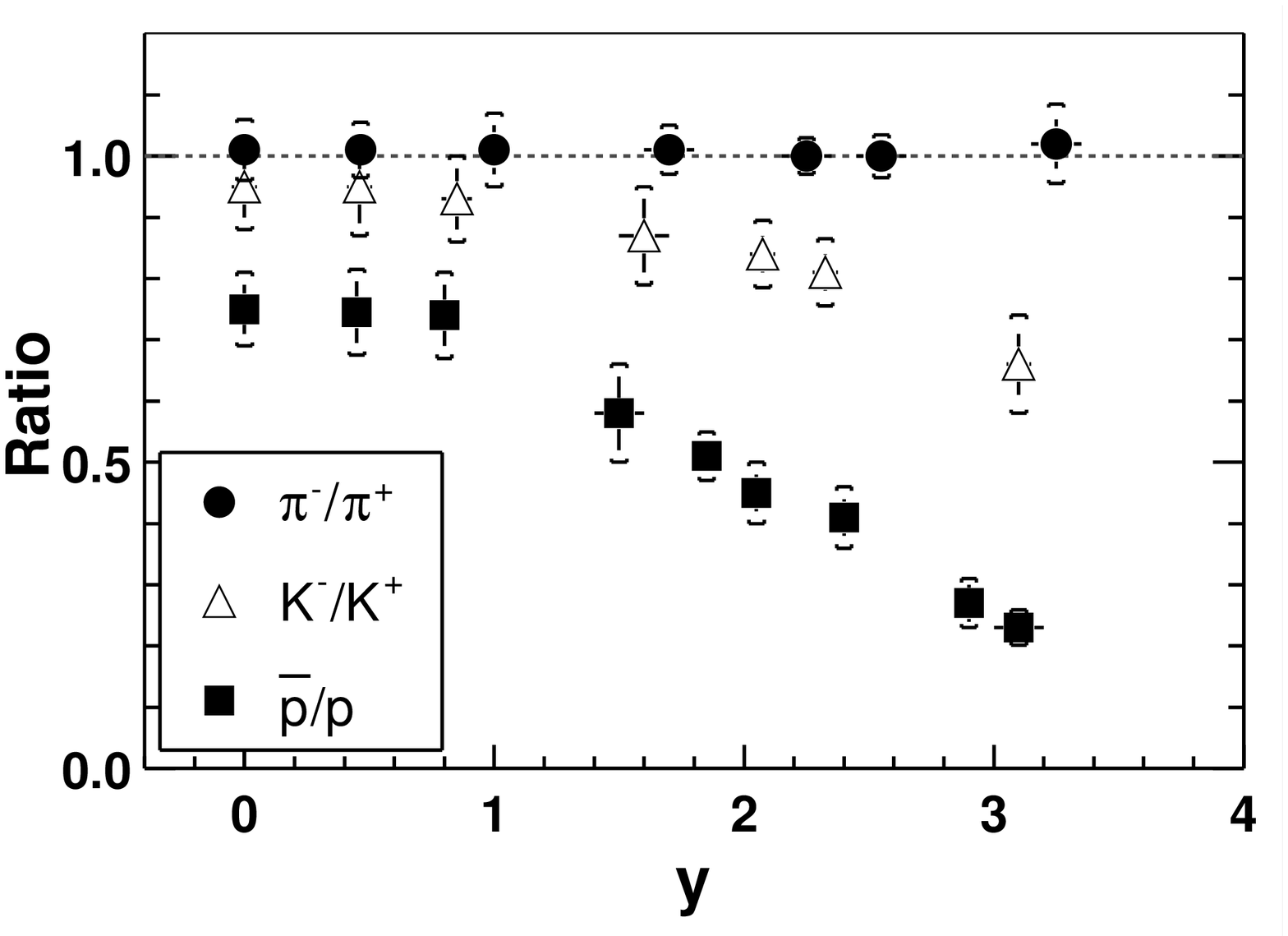}
\caption{\it Ratios of particles to antiparticles vs. rapidity, from the
BRAHMS experiment.}
\label{fig:brahms-p-stop.eps}
\end{minipage}
\hspace{\fill}
\begin{minipage}[t]{75mm}
\raisebox{.5cm}
{
\includegraphics[width=75mm]{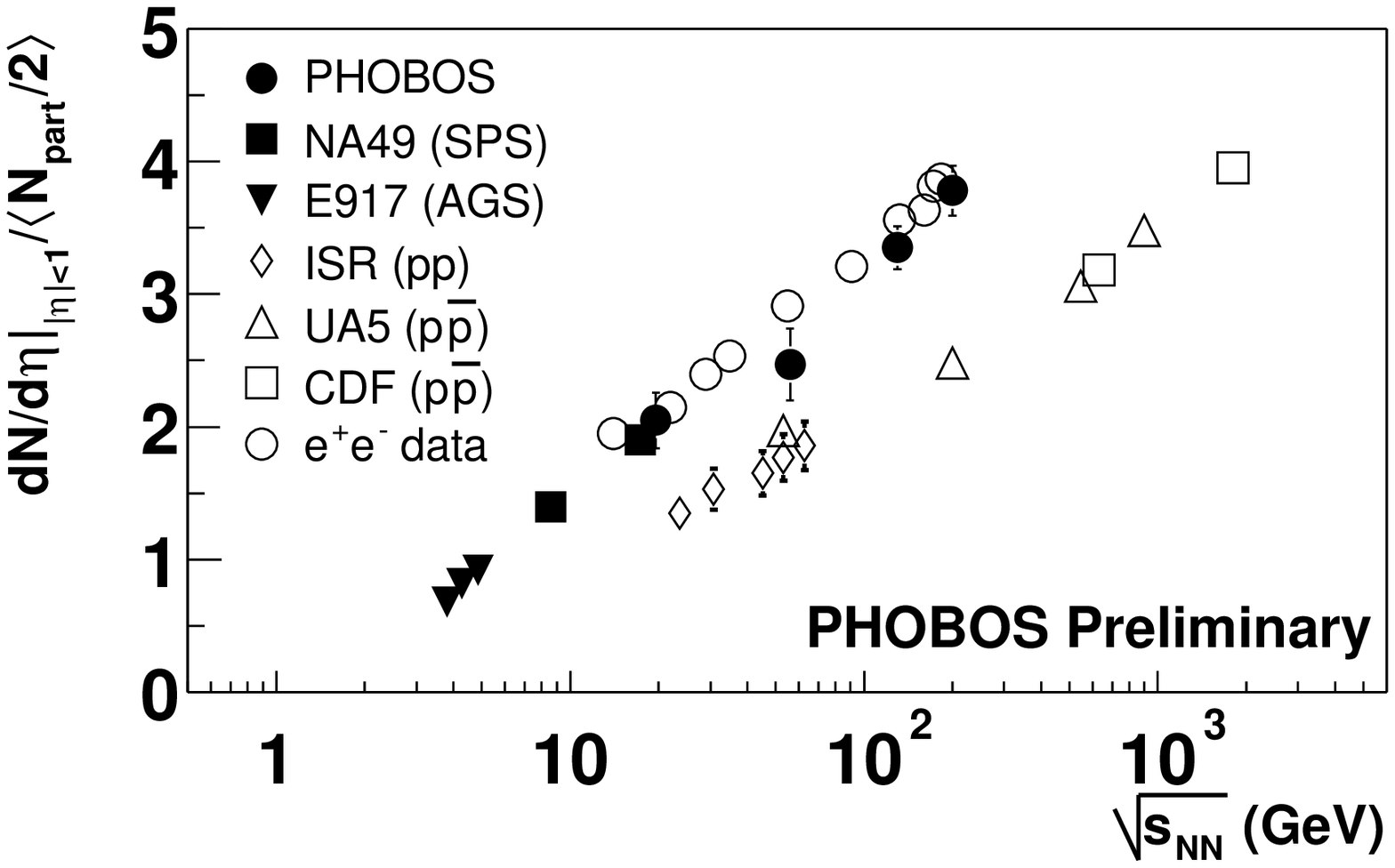}
}
\caption{\it Charged particle density at midrapidity, for $\epem$, $\pp$
and Au+Au collisions.}
\label{fig:midrap_AA_ee_pp}
\end{minipage}
\end{figure}

The initial conditions of heavy ion collisions are determined 
primarily by the beam energy and the nuclear geometry.
The impact parameter between the nuclei 
controls the overlap area of the contracted nuclei and thus 
the number of nucleons which participate via inelastic processes
($\np$) and the number of binary collisions ($\nc\propto\np^{\frac{4}{3}}$)
The beam energy determines the initial state parton density, 
which is dominated by gluons at low-$x$, the available
range in rapidity ($\Delta y = 2\ln(\s/m_p)$), 
and the total nucleon-nucleon cross section.
The combination of these factors
contribute to the amount of energy deposited by the incoming
baryons in each nucleus as they lose energy while penetrating
the other nucleus -- a phenomenon called ``proton stopping''
This is typically measured by the net rapidity distribution of protons
minus anti-protons.  Recent results from BRAHMS
\cite{brahms-stopping} show that
while the ratio of protons to anti-protons is about 0.75 at 
mid-rapidity, it falls rapidly by $y=3$, consistent with a
rapid increase in the net proton density, which is expected to
peak at a rapidity $2.5$ units from $y_{\rm beam}$\cite{stopping}.
These measurements will be
critical for understanding the very early collision stage.

All of the RHIC experiments perform measurement
at midrapidity,
where hard processes are expected to contribute substantially
to particle production.  
The energy density achieved in the collisions can be estimated
either by measuring the transverse energy and using the Bjorken
formula $\epsilon = (dE_{T}/dy) / \pi R^2 \tau_o$, 
with $R$ the nuclear radius and $\tau_o$ the ``formation time''
of the system.
PHENIX has measured the transverse energy
density at midrapidity in 130 GeV Au+Au collisions
and extracted an energy density
$\epsilon =$ 4.6 $\gevfm$ \cite{phenix-et} assuming at
typical value for $\tau_o \sim 1$ fm/c.  This is already well above
lattice expectations for a QGP phase transition. 
They have also found the ratio $E_T/N_{ch}$ to be constant for all $N_{ch}$.
Thus, one can use the ratio of midrapidity charged particle production at 
130 and 200 GeV ($R=1.14\pm.05$), 
shown in Fig. \ref{fig:midrap_AA_ee_pp} as
the particle production per participant pair \cite{phobos}, 
to infer an energy density of about $\epsilon = 5.3$ $\gevfm$ at 
the highest RHIC energies.

\begin{figure}[h]
\begin{minipage}[t]{75mm}
\includegraphics[width=75mm]{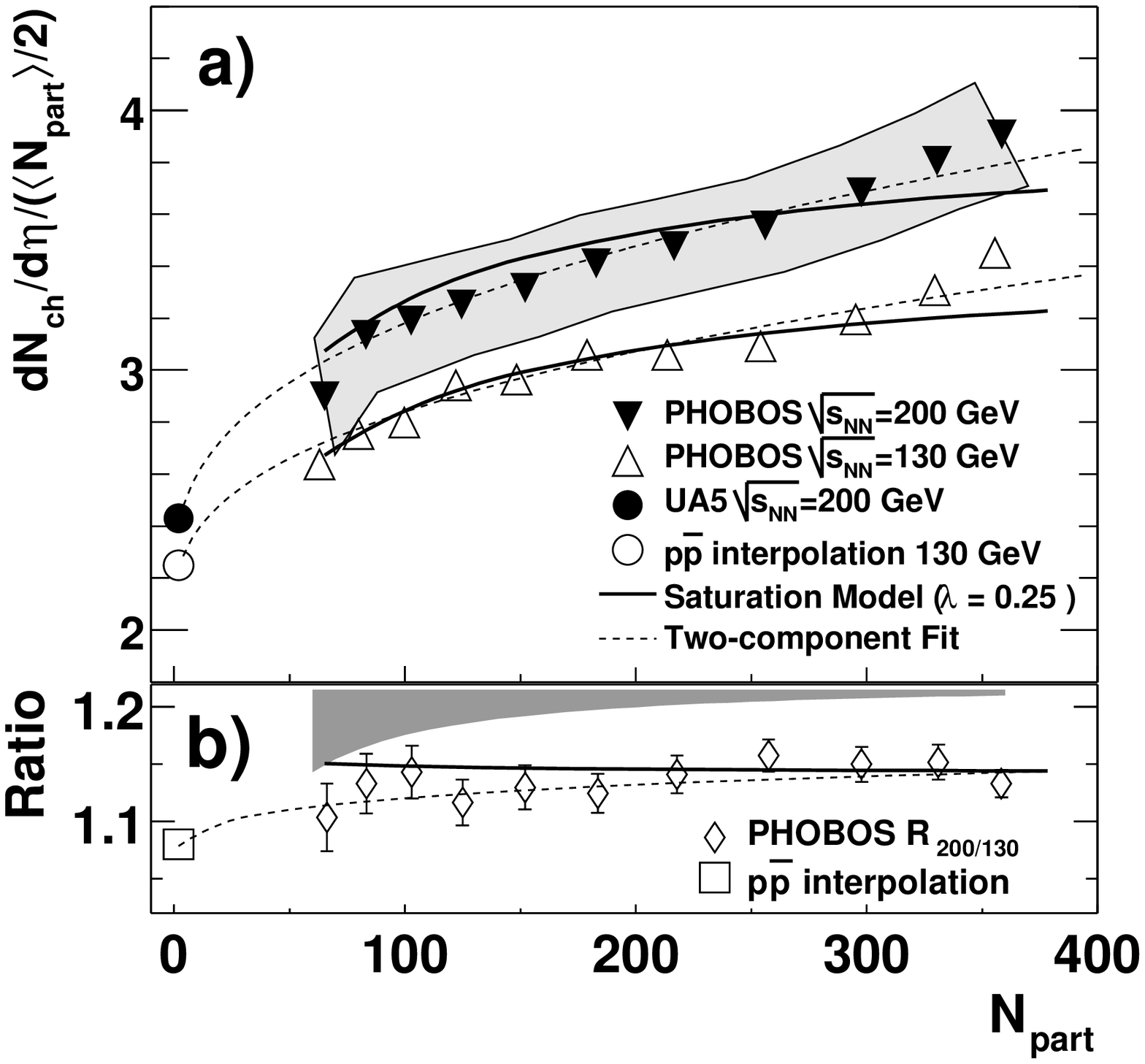}
\caption{\it Centrality dependence of $\dndetanp$ compared
with two-component and saturation models.}
\label{fig:final_tracklet_200}
\end{minipage}
\hspace{\fill}
\begin{minipage}[t]{70mm}
\includegraphics[width=70mm]{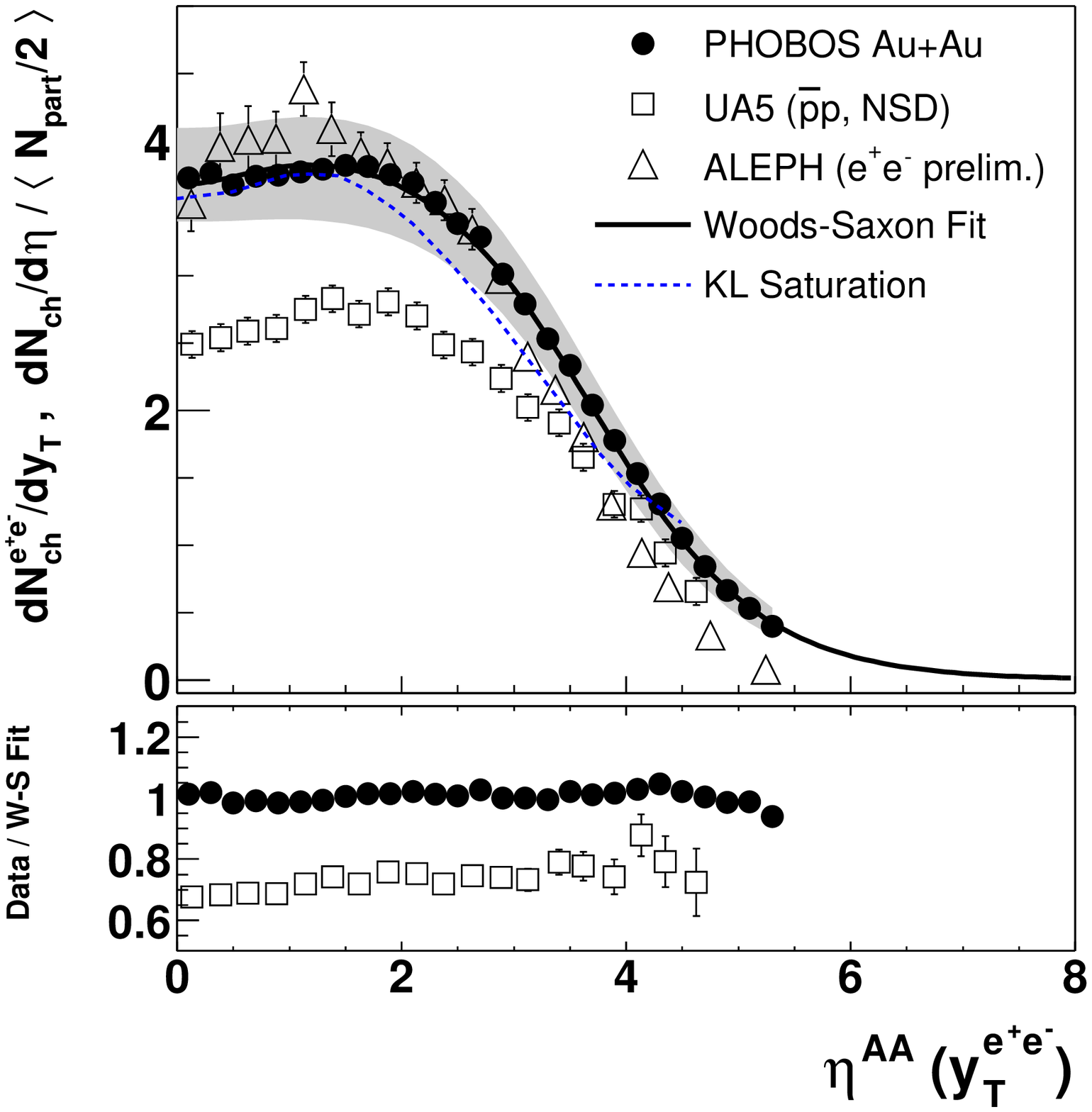}
\caption{\it Pseudorapidity distributions of Au+Au, $\pbarp$
and $\epem$ at $\s$=200 GeV.  The parton saturation
prediction of Ref. \cite{kharzeev-levin} is also shown.}
\label{fig:AA_ee_pp_kln}
\end{minipage}
\end{figure}

The centrality dependence of charged particle production,
$\dndetanp$, offers a means to study the evolution
between $\pp$ collisions and central Au+Au.  Two types of 
predictions \cite{kharzeev-nardi} have proven effective in 
describing the midrapidity data \cite{phobos} shown in Fig. 5.
The ``two-component'' model postulates that the total
particle production stems from a linear combination of 
soft processes that scale with the number of participants,
and hard processes that scale with the number of binary collisions.
Thus, $\dndeta = \npp((1-X(s))\np/2 + X(s)\nc)$, where $\npp$ is the
measured multiplicity in $\pp$ collisions, and $X(s)$ is the
energy-dependent fraction of hard processes in $pp$ collisions.
(approximately 0.1 at RHIC).
It is also expected that the energy released in the form of
low-$x$ gluons should create a parton density so high 
that the partons below a ``saturation scale'' $Q_s$ (i.e. of larger
transverse size) recombine in order not to exceed a maximum
value of order $1/\as$ \cite{saturation}. 
In this picture, particle production is determined mainly by $Q_s$, 
which itself depends on the transverse density of
partons (i.e. $Q^2_s \propto A^{1/3}$), $\dndetanp = 1/\alpha_s(Q^2_s)$.
As shown in Fig. \ref{fig:final_tracklet_200}, both of these
models offer an efficient description of the experimental
data from all of the RHIC experiments.
Saturation models are also able to offer a reasonable
quantitative description of the full rapidity distribution,
an example of which by Kharzeev and Levin \cite{kharzeev-levin}
is shown as a dotted line in Fig.
\ref{fig:AA_ee_pp_kln}.  

The success of the saturation models
in capturing basic features of particle production has
a fascinating consequence.  While most expectations of the
``formation time'' $\tau_0$ used in energy density
estimates are around 1 fm/c, saturation models predict
the formation time of the initial gluon state to be
$\tau_s \sim \hbar/Q_s \sim .2$ ${\rm fm/c}$, which
gives $\epsilon \sim 18$ $\gevfm$ when plugged into the
Bjorken formula for the 130 GeV data.
While neither estimate (5 $\gevfm$ or 18 $\gevfm$) should be taken
too seriously at this point,
these estimates show that both naive and more sophisticated
estimates arrive at values well above the lattice 
expectations for QGP formation.

\section{Universal features of particle production}

\begin{figure}[tb]
\begin{minipage}[t]{70mm}
\includegraphics[width=7cm]{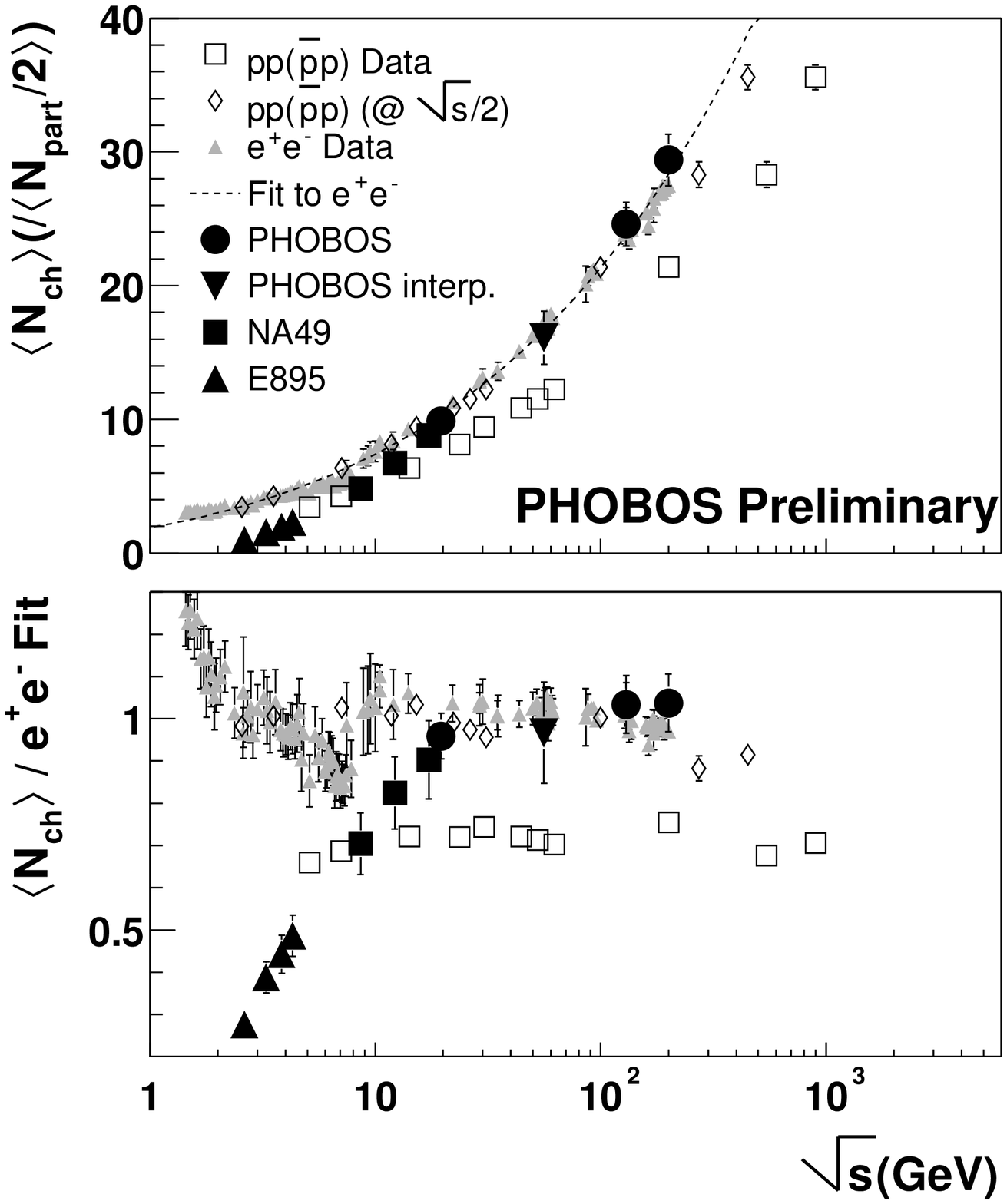}
\caption{\it Total charged particle production for a
variety of strongly-interacting systems.  Comparison
with a fit to the $\epem$ data is shown.}
\label{fig:total_ratio}
\end{minipage}
\hspace{\fill}
\begin{minipage}[t]{75mm}
\includegraphics[width=75mm]{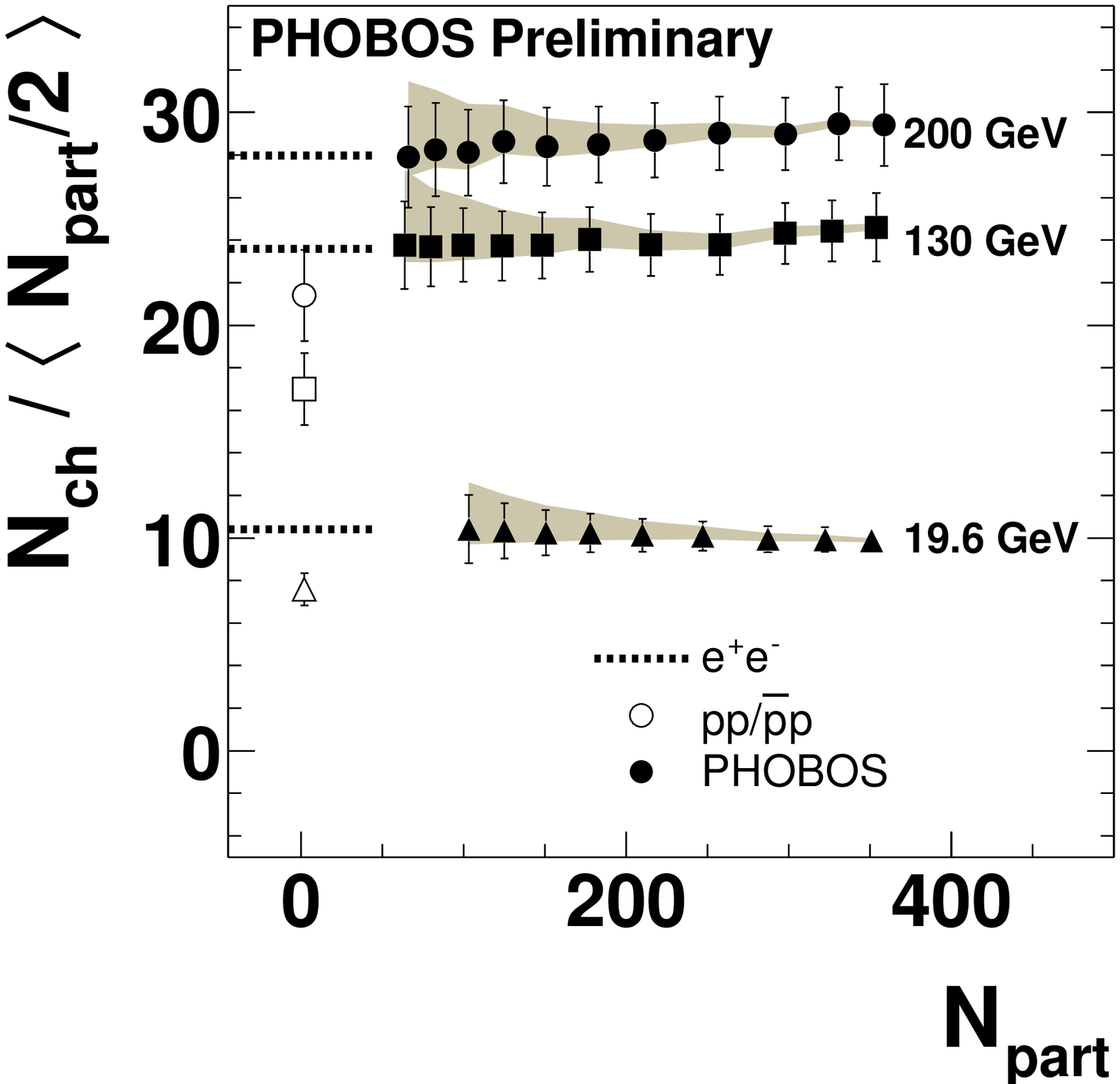}
\caption{\it Total multiplicity per participant pair
for Au+Au collisions at $\snn=19.6, 130$, and $200$ GeV.}
\label{fig:ntot_20_130_200}
\end{minipage}
\end{figure}

We can gain insight into
the relationship between Au+Au collisions and elementary
nucleon-nucleon interactions by comparisons with $\epem$
annihilations to hadrons.
It has long been noted that the charged multiplicities
in $\pp$ and $\epem$ collisions are similar if the
energy taken away by the ``leading baryons'' is subtracted
from $\s$ and then compared to $\epem$ collisions at this
$\seff$, suggesting a universal fragmentation
mechanism \cite{basile}.
With new data from RHIC and LEP2, PHOBOS finds
that Au+Au (divided by the number of participant pairs
$\halfnp$) and $\epem$ collisions have similar pseudorapidity 
densities over a large range (see Fig. \ref{fig:AA_ee_pp_kln})
and thus a similar total multiplicity \cite{phobos-universality}.
Comparisons of the total multiplicity per participant
pair $\ratio$ over a large range
of CMS energies is shown in Fig. \ref{fig:total_ratio}
and compared to $\epem$ and $\pp/\pbarp$ data.
It is easiest to compare these data sets by dividing
all of the values vs. $\s$
by a function that describes the $\epem$ data \cite{mueller}.
It is seen that the Au+Au and $\epem$ data converge for 
$\s > 20$ GeV, while the $\pp$ data also follows the same trend
if the effective energy $\seff = \s/2$ is used.

The total multiplicity produced per participant pair
in Au+Au collisions has also been measured by PHOBOS.
It is shown in Fig. \ref{fig:ntot_20_130_200}
to be constant over the range of centralities
measured ($\np>65$) for all three energies
\cite{phobos-universality}.
This is reminiscent of ``wounded-nucleon'' scaling 
\cite{elias} but 
in all three cases the multiplicity is about 30-40\% higher
than $\pp$ at comparable energy, and comparable to $\epem$.  
This adds an interesting perspective to the
use of minijet and saturation approaches in describing
the multiplicity at $\eta=0$, as discussed in the previous section.

This similarity in bulk particle production between
Au+Au and $\epem$ data for $\s>20$ GeV
suggests that Au+Au are more efficient
than $NN$ interactions in 
transferring the incoming energy into particle production.
This might simply indicate that the multiple collisions
suffered per participant substantially reduce the 
leading particle effect.
The drop below 20 GeV may be explained by the presence
of a large baryon density in the final state \cite{na49}.
While it is obvious that Au+Au collisions do not start
with the same initial scale as $\epem$, which is reflected
in the much harder $\pt$ distributions in $\epem$ above 2 GeV, 
it appears that the available energy, rather than
the initial hard scale, has the stronger effect on the
bulk particle production.  This may be another
manifestation of the universality of low-$x$
physics in strongly-interacting systems at high
energy (which is discussed in the HERA context in Ref.
\cite{devenish}).  
The physics behind this may need to be addressed before
claims to particular behavior for particle yields in
heavy ions can be made in other contexts.

\section{Dynamics of the Parton Cascade}

\begin{figure}[tb]
\begin{minipage}{75mm}
\includegraphics[width=75mm]{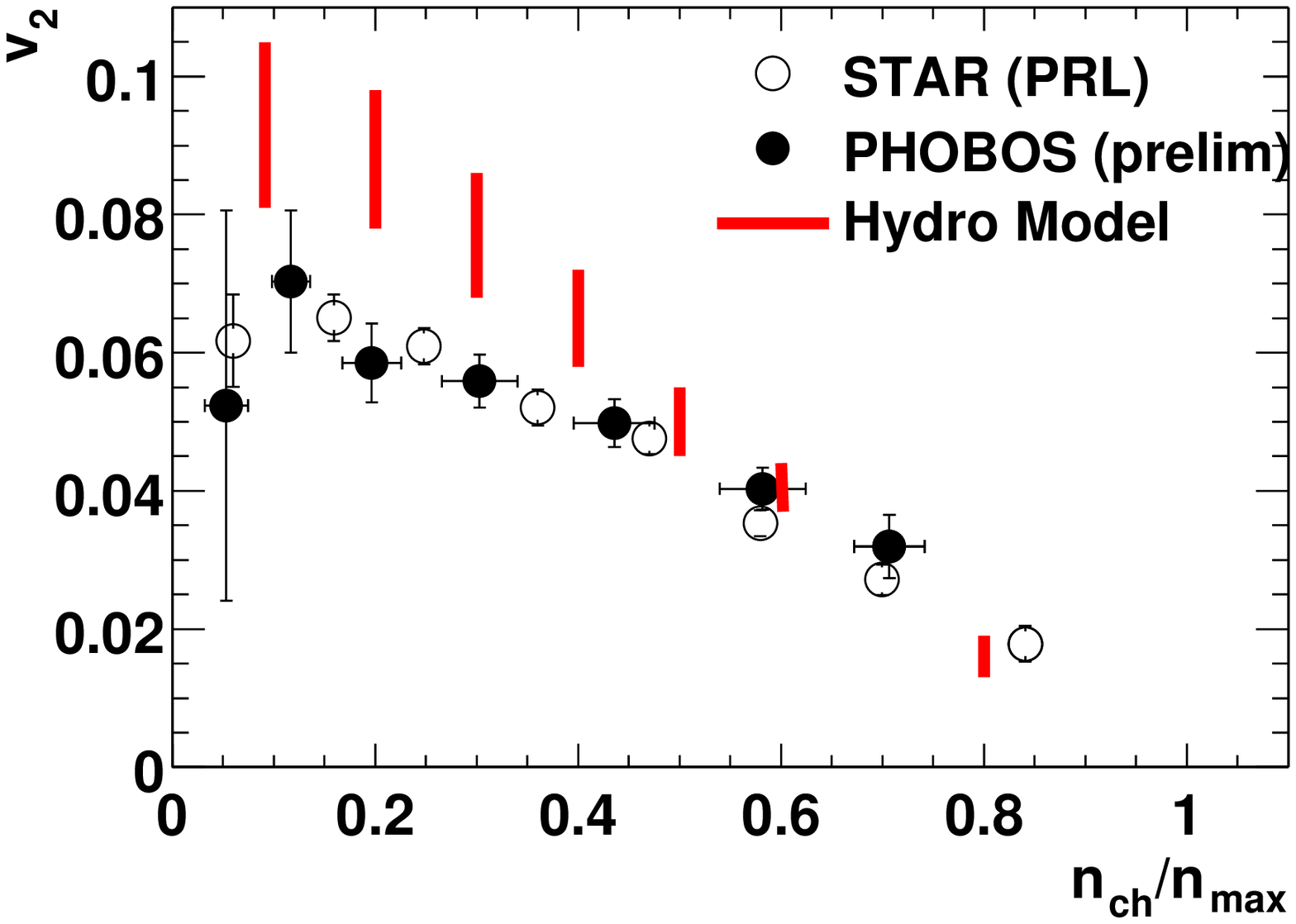}
\caption{\it Data from STAR and PHOBOS on
elliptic flow vs. collision centrality compared
to hydrodynamic predictions.
\label{fig:all_flow2}}
\end{minipage}
\hspace{\fill}
\begin{minipage}{70mm}
\includegraphics[width=75mm]{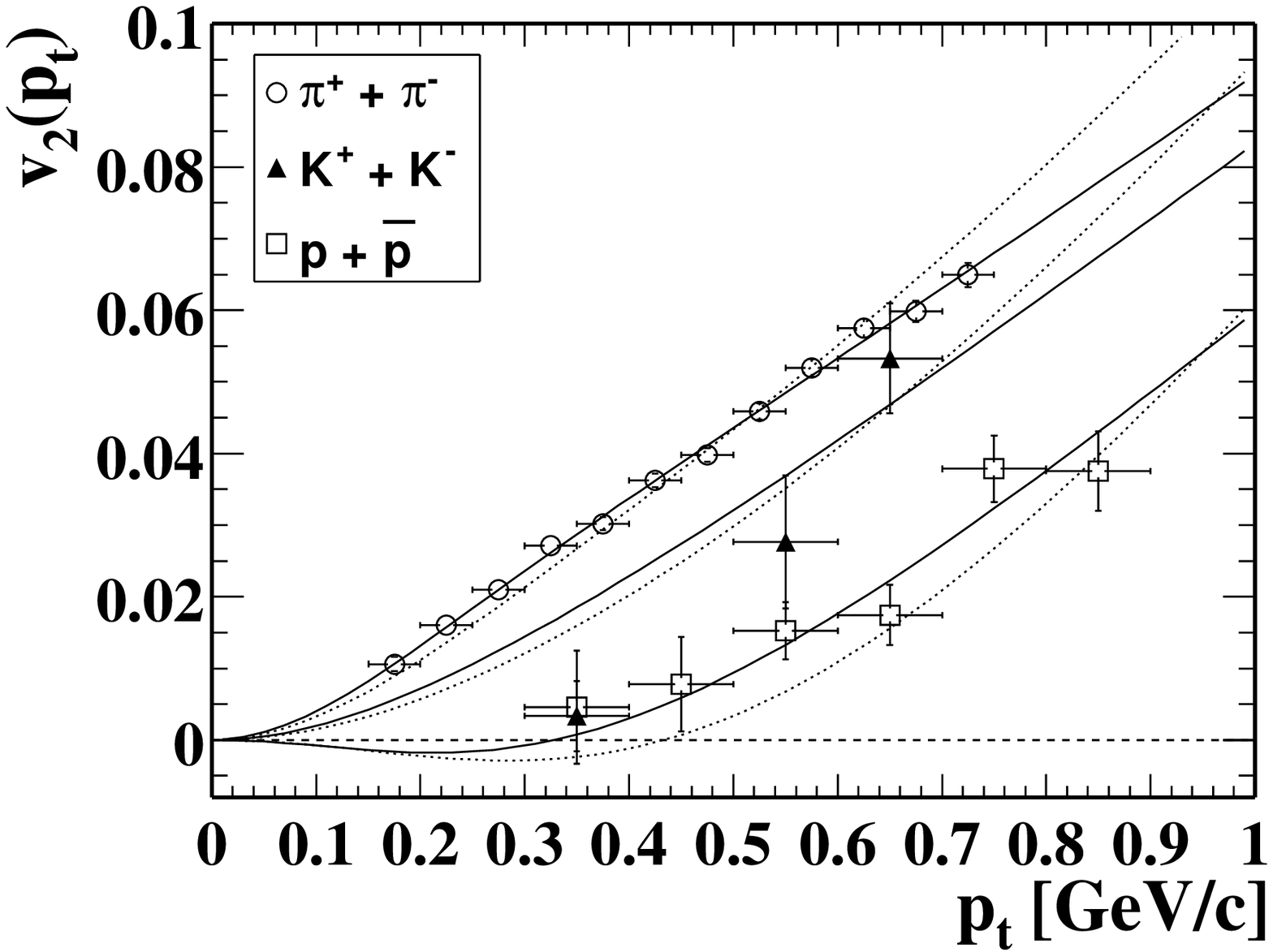}
\caption{\it STAR data on $\vt$ vs. $\pt$ for different
particles species, also compared with hydro calculations.
\label{fig:flow-vs-pt}}
\end{minipage}
\end{figure}

If the energy density is indeed as large, and forms
as early, as the success of the saturation descriptions
seem to imply, then we might expect the density of
scatterings to lead to an early thermalization and
possibly the development of hydrodynamic behavior
as the system evolves \cite{heinz-kolb}.
This expectation is borne
out by measurements of the event-by-event azimuthal
distributions, which are found to have a slightly
elliptical shape at all energies from AGS to RHIC.
The maximum value of $\vt$, the second Fourier component
of the azimuthal distribution,
$dN/d\phi \propto 1 + 2 \vt \times \cos(2(\phi-\Psi_R))$
with $\Psi_R$ the angle of the event reaction plane,
has been found to increase with increasing beam
energy.  This is contrary to what one would expect
if the dynamics were purely hadronic, since late
stage rescattering would wash out any initial state
anisotropy.  Thus, the non-zero $\vt$ has been
interpreted as stemming from hydrodynamic pressure gradients
in the initial state, implying partonic thermalization,
which is an essential prerequisite to QGP formation.

The results from STAR and PHOBOS \cite{star-flow, phobos-flow}
are shown in Fig. \ref{fig:all_flow2} and compared with
predictions from a range of hydro models (shown by
solid bars).  These results suggest that
the more central events at RHIC are in fact consistent
with hydrodynamic evolution.  The models used here
typically start with energy densities around 20 GeV/fm$^3$
evolving from an early time $\tau < 1$ fm,
which is consistent with values extracted from the saturation
approach.

The importance of $\vt$ as an experimental observable comes
from the quantitative comparisons with theoretical predictions
for observables relative to the reaction plane, which is
defined as the axis where $dN/d\phi$ is at a maximum.
Measurements of $\vt$ as a function
of the transverse momentum \cite{star-flow} characterize
the modulation of $dN/d\phi$ for particles of a given momentum
relative to the reaction plane.
As seen in Fig. \ref{fig:flow-vs-pt}, STAR has measured
$\vt(\pt)$ for different particle species and found
good agreement with hydro calculations out to $\pt\sim 1$ GeV,
where their particle identification stops.  However, two
notable disagreements with hydro exist.  The linear rise of
$\vt$ with $\pt$ stops at $\pt\sim 2$ GeV 
and remains constant out to larger $\pt$ \cite{star-flow}.
Also, the pseudorapidity distribution of $\vt$ drops
rapidly away from midrapidity, suggesting that the ``boost-invariant''
hydrodynamic behavior may only be found very near $y=0$ \cite{phobos-flow}.

\section{Thermal Description of the final Hadronic State}

As the system expands and cools, it eventually hadronizes into
the final state particles measured in the detectors.
As these are the particles identified by most of the experiments,
a large amount of data exists on the properties of this stage,
only a subset of which will be shown in this section.
The most pressing issue is whether
the final state shows collective effects, which would be
a signal of equilibration, as opposed to particle production
simply proceeding via the available phase space.

\begin{figure}[tb]
\begin{minipage}[t]{60mm}
\raisebox{5mm}
{
\includegraphics[width=60mm]{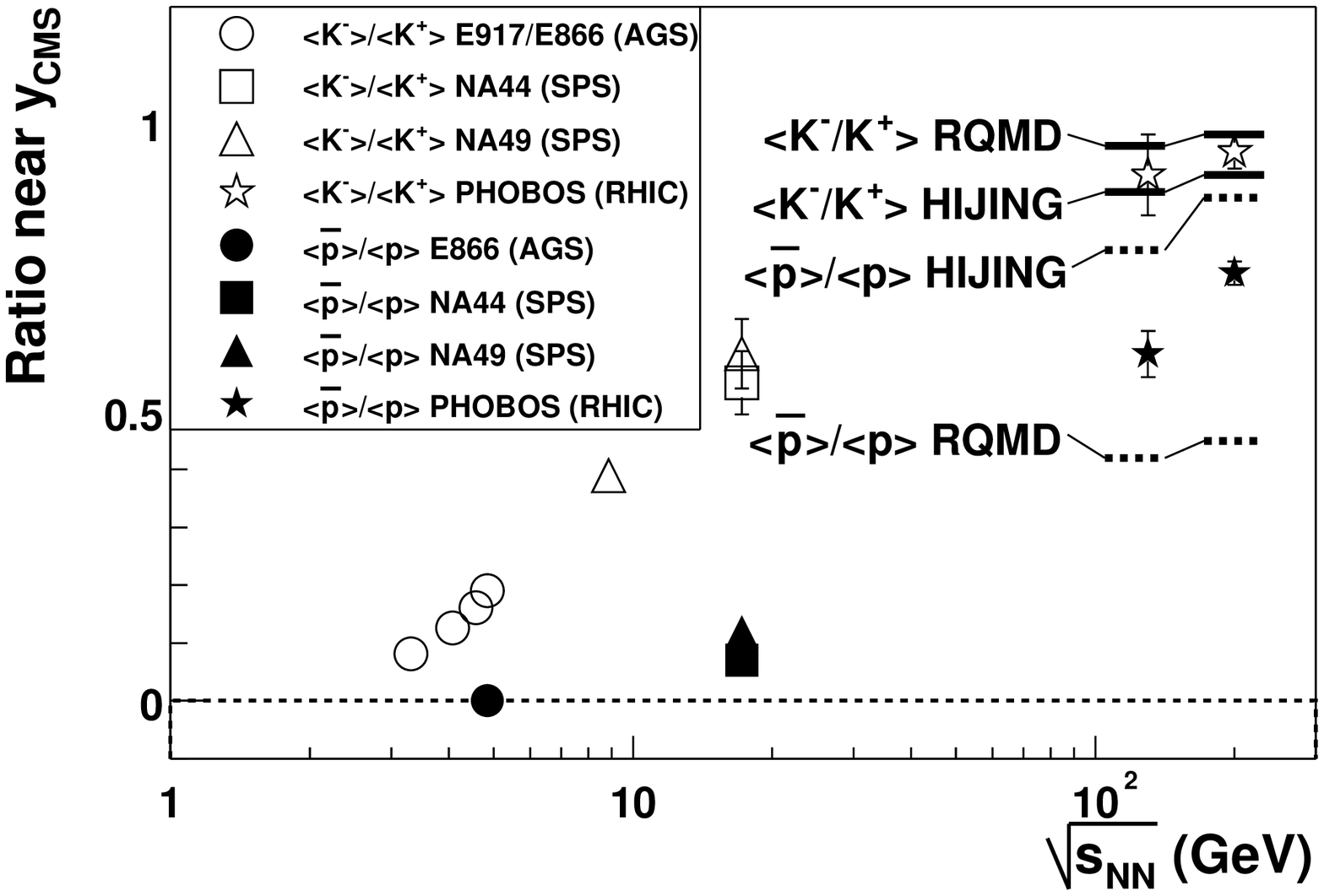}
}
\caption{\it PHOBOS ratios of anti-particle to particle vs. $\s$}
\label{fig:particle-ratios}
\end{minipage}
\hspace{\fill}
\begin{minipage}[t]{90mm}
\includegraphics[width=90mm]{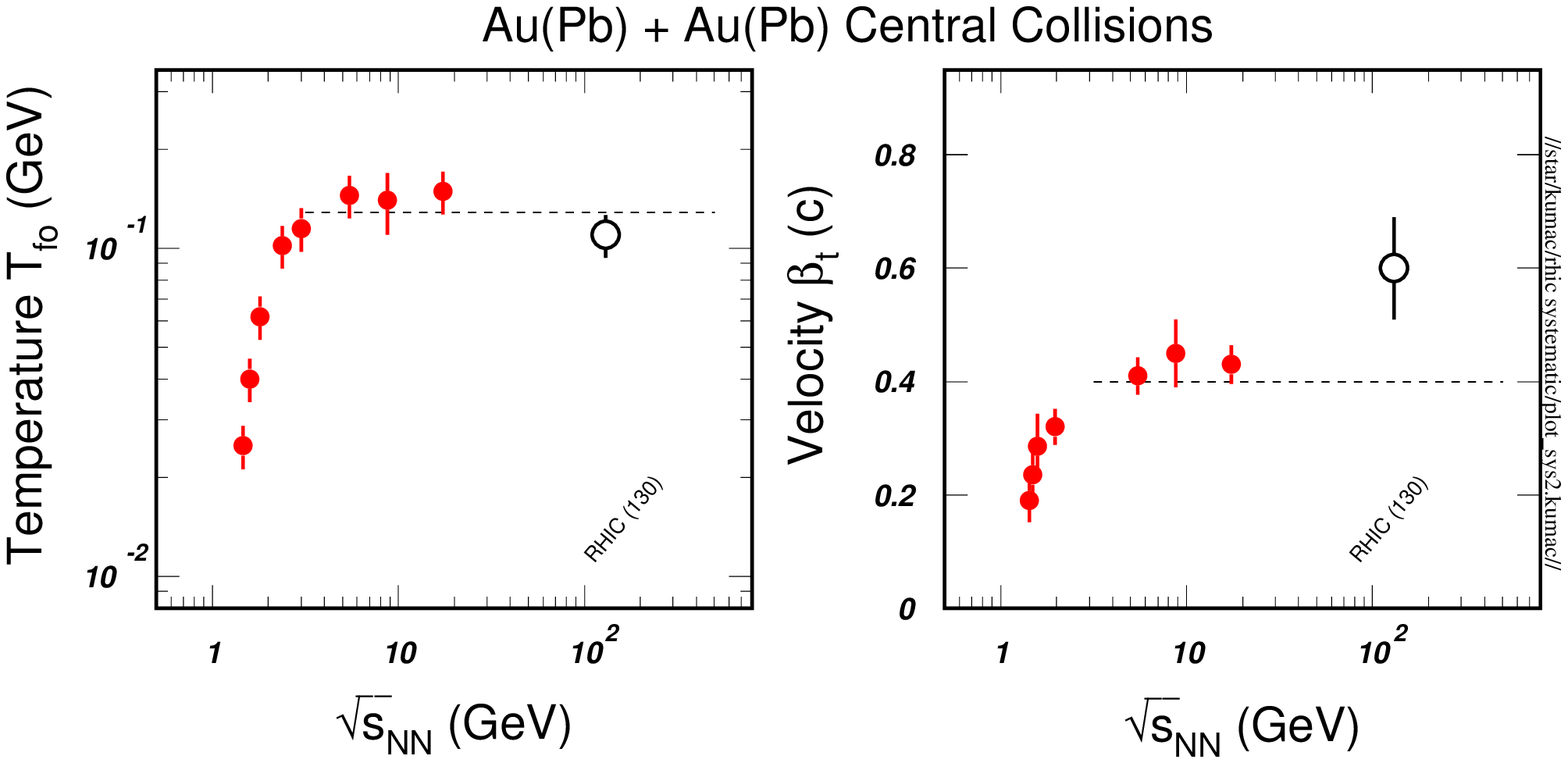}
\caption{\it Thermal freezeout parameters shown as a function
of $\s$.
\label{fig:plot_sys2}
}
\end{minipage}
\end{figure}

The most basic property of the system, measured by all
four RHIC experiments, is the gradual approach of all
of the anti-particle/particle ratios towards unity \cite{phobos-ratios},
as shown in Fig. \ref{fig:particle-ratios}.  
The fact that $\overline{p}/p$ is less than unity 
directly indicates the presence
of some fraction the primary baryons at mid-rapidity,
an incredible fact considering the large initial baryon momentum.
The kaons, while approaching one, also do not reach it
at the top RHIC energy. 
These two ratios have been found
to be highly correlated, suggesting that the produced
hadrons are highly sensitive to the net baryon density.
Still, RHIC is closer than ever before to creating the conditions
found in the very early universe.

The relationship between these ratios, and many others,
can be understood by means of ``thermal models''\cite{thermal-models}, 
which assume the system to be in chemical and thermal equilibrium
up to the time when the momentum transfers are low enough that
the flavor composition of the system is frozen.
If only the ratios of particle yields are used, then
the available parameters are simply
$T$ (the temperature), $\mu_B$ (the baryon chemical potential) 
and $\gamma_s$, which parametrizes
the deviation from full strangeness equilibration.  
A volume ($V$) is needed if absolute yields are
to be characterized as well.

The application of these models to data from central
collisions of heavy nuclei, as a function of $\s$, finds the
simple result (shown in Fig. \ref{fig:qgp-phase-diag}) that all of
the data seem to lie on a single contour on the $( T , \mu_B )$
space.  There are two interpretations of this behavior \cite{thermal-models}.
Cleymans {\it et al.} explain this as freezeout occurring when 
the energy per particle  $\langle E \rangle
/ \langle N \rangle \sim$ 1 GeV.  
Braun-Munzinger and Stachel suggest that
it occurs at a constant total baryon density of 
$n_B = .12 / {\rm fm}^3$.  Both of these interpretations
are consistent with the existing data.

It is interesting that thermal models have also been applied
to $\pp$, $\pbarp$ and $\epem$ data and find the similar
result of $T \sim 170$ MeV (with $\mu_B = 0$) \cite{becattini}.
This suggests that the heavy ion data above CERN energies 
approaches similar freezeout state as that for elementary systems.
This may be another indication of the onset of universal
behavior as the role of the net baryon density becomes less
significant.  However, some important remain between
heavy ions and the elementary systems.

One difference between nuclei and the more elementary
systems is evident in the details of the momentum spectra.
In elementary collisions ($\pp$, $\epem$), the spectra
at low $m_T$ show ``$m_T$ scaling'' where the individual particle
yields described by an exponential $\exp(-m_T/T)$ with
same slope for all particle species.  In heavy ions, 
the slope at low $m_T$ increases with the particle mass
in a way which can be fit by the form $T_{\rm eff} = 
T_o + m \langle \beta \rangle^2$, where 
$\langle \beta \rangle$ is a collective ``radial flow'' velocity.
The energy dependence of the fit parameters
has been extracted by Kaneta and Xu \cite{kaneta-xu} and 
is shown in Fig. \ref{fig:plot_sys2}.   They observe a 
thermal freezeout temperature ($T_{th}=140$ MeV) 
which is consistently lower than the
chemical freezeout extracted from the particle ratios
($T_{ch}=170$ MeV).  They
also find that $\langle \beta \rangle$ increases rapidly wity $\s$
but appears to saturate above $\snn \sim 10$ GeV.  
These data strongly imply collective behavior among the final
state hadrons.

\begin{figure}[t]
\begin{minipage}{70mm}
\includegraphics[width=70mm]{jaipur2_fig4.eps}
\caption{
\it Energy and system dependence of the Wroblewski factor, which
measures the ratio of strange to non-strange particle
production.
}
\label{fig:jaipur2_fig4}
\end{minipage}
\begin{minipage}{75mm}
\begin{center}
\includegraphics[height=75mm]{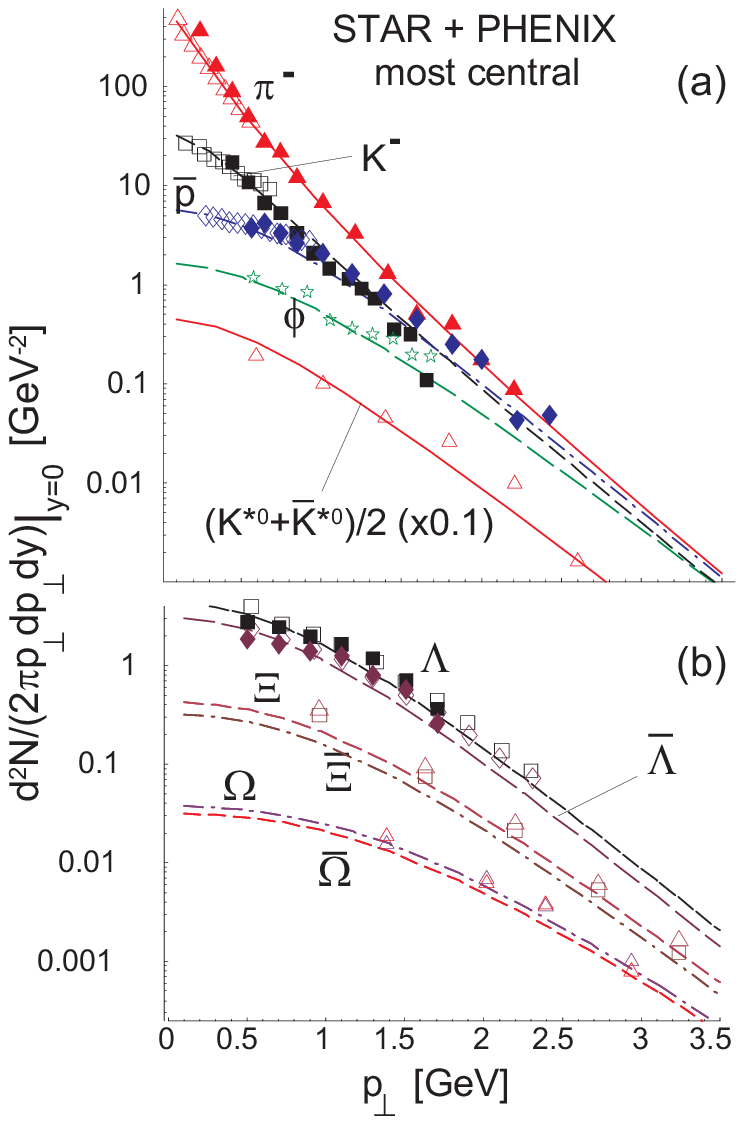}
\end{center}
 \caption{\it 
Single freeze-out time calculations for
identified particle
spectra by Broniowski and Florkowski.
    \label{fig:bron-flor}}
\end{minipage}
\end{figure}

Another important difference is seen in the total strangeness 
yield, which 
is ``enhanced'' by a factor of two in heavy ion collisions, as shown by
the energy dependence of the ``Wroblewski'' factor
($\lambda_s = 2\langle\overline{s}s\rangle/
(\langle\overline{u}u\rangle+\langle\overline{d}d\rangle)$) in Fig.
\ref{fig:jaipur2_fig4}.  While strangeness enhancement has long
been a canonical signal of QGP formation,
there exists a simple interpretation of this effect in terms 
of the volume over
which strangeness can equilibrate.  In other words,
the system produced in nuclear collisions no longer
seems to be forced to obey {\it local} strangeness
conservation \cite{strangeness}.
However, a recent analysis by NA49 \cite{hohne},
characterizing the $K^+/\pi^+$ ratios as a function of centrality
at CERN SPS energies, has shown that the effect may not
be related precisely to the reaction volume.  Instead,
this ratio scales with the fraction of participants that have
collided more than once, shifting the responsibility for
what appears to be equilibration in the final state to
some property of the initial collision geometry.

Both chemical and thermal freezeout is addressed in the 
the model by Broniowski and Florkowski \cite{bronflor}. 
They assume that chemical and
thermal freezeout occur at the {\it same} time.
By fitting to the available $\pi$, $K$ and $p$ spectra
from PHENIX and STAR,
and incorporating transverse expansion, this model is
able to describe all measured particle species, and was
even successful in predicting particles like the $\kstar$
and $\Omega$.  The success in understanding the  $\kstar$ yields
is very interesting since it has a lifetime comparable to the lifetime of the
system itself ($c\tau \sim 4$ fm).  This suggests that there
is little, if any, subsequent interactions of the $\kstar$ decay products, 
lending credibility to the assumption of the single freezeout time.
This model is similar in concept to so-called ``blast wave'' 
fits to experimental data, which also assumes that hadronization
occurs on a thin space-time shell.  The success of these two
approaches suggests that the hadronization process is quite sudden.

\section{Hard Probes of the Early Stages}

Hard processes occur at early times ($\Delta t \sim \hbar/\Delta E$) and thus
offer a means to probe the early stages of the collision.
In principle, jet cross sections can still be calculated within
a pQCD framework as they are for hadron-hadron interactions,
by considering the structure of the nucleons in the colliding
nuclei and assuming they interact via the standard pQCD matrix 
elements.  Naively, one would expect that the rate of hard
processes should scale with the number of binary collisions,
since the timescale of hard processes is so short that 
incoming nucleons will be sensitive to all of the nucleons
in its path.  However, there is large body of theoretical
work suggesting that high-energy partons should lose energy
in a deconfined medium via color brehmsstrahlung -
a phenomenon called
``jet quenching''.  Jets traversing a hadronic
medium should suffer no induced radiation and thus be
unmodified.


At RHIC energies, there are two impediments to directly measuring
jets.  First of all, the typical energy is not high enough 
to create well-collimated cones of particles.  
Secondly, a central event always
creates a large background of soft particles which will obscure a 
jet signal within a typical jet cone.  Thus, measurements of quenching
focus primarily on the high-$\pt$ part of the inclusive hadron spectrum,
to look for modifications to the measured jet fragmentation functions.

First results on particle spectra at high $\pt$ in heavy ion
collisions were shown by the PHENIX and STAR collaborations only
a few months after the end of the 130 GeV RHIC running in 2000
\cite{high-pt}.
Both collaborations presented the modification of the 
inclusive hadron spectra (both of all charged particles and 
identified $\pi^o$'s) in central Au+Au collisions by a ratio 
comparing it with
spectra from $\pbarp$ collisions measured by the UA1 collaboration,
interpolated to 130 GeV and then divided by the number of binary
collisions.

\begin{figure}[t]
\begin{minipage}{70mm}
\centerline{\epsfig{file=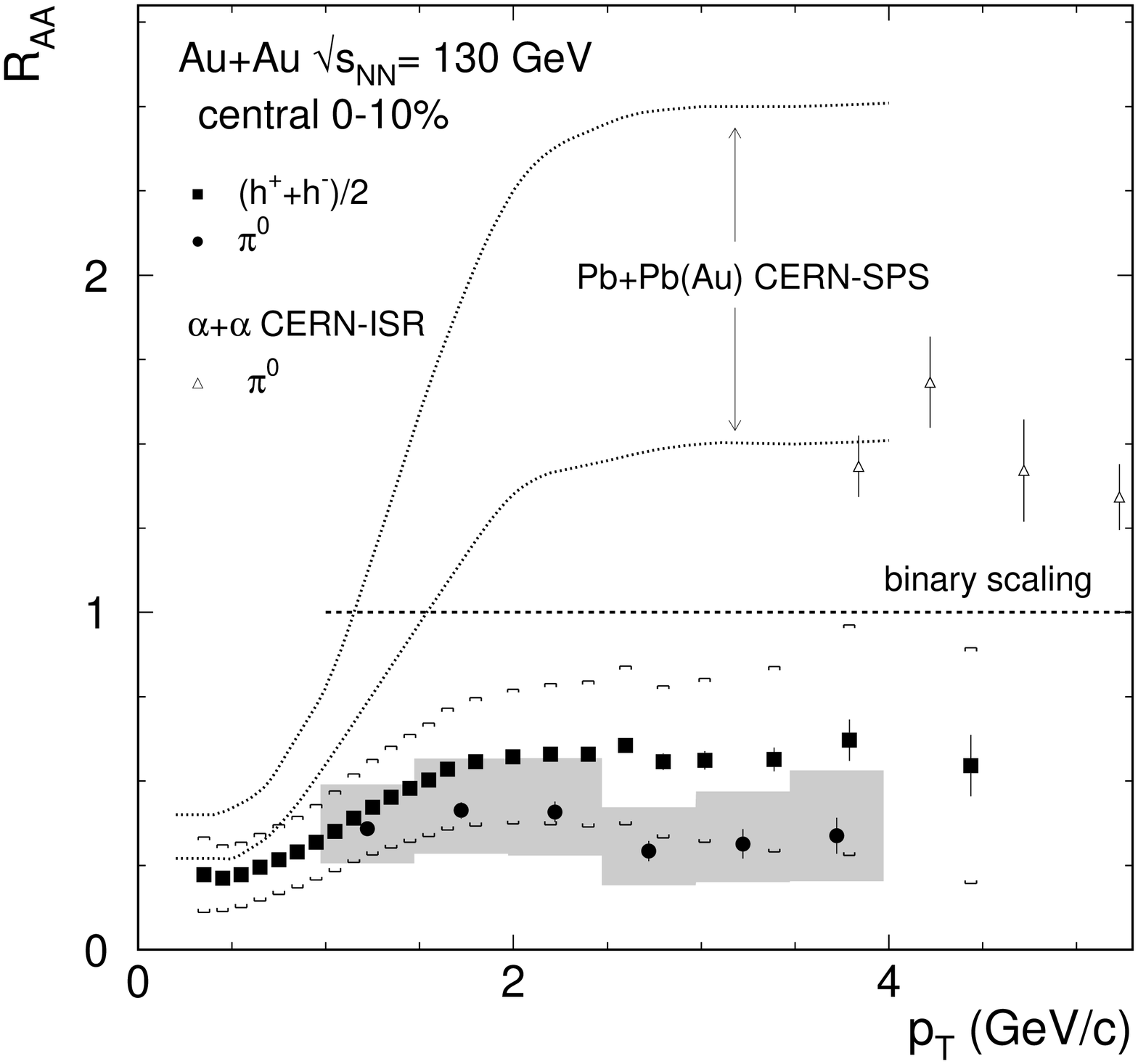, width=75mm}}
 \caption{\it 
PHENIX results indicating suppression of high-$\pt$
particle production in nuclear collisions.
 \vspace{0.5cm}
    \label{fig:phenix-jetq}}
\end{minipage}
\begin{minipage}[r]{75mm}
\centerline{\epsfig{file=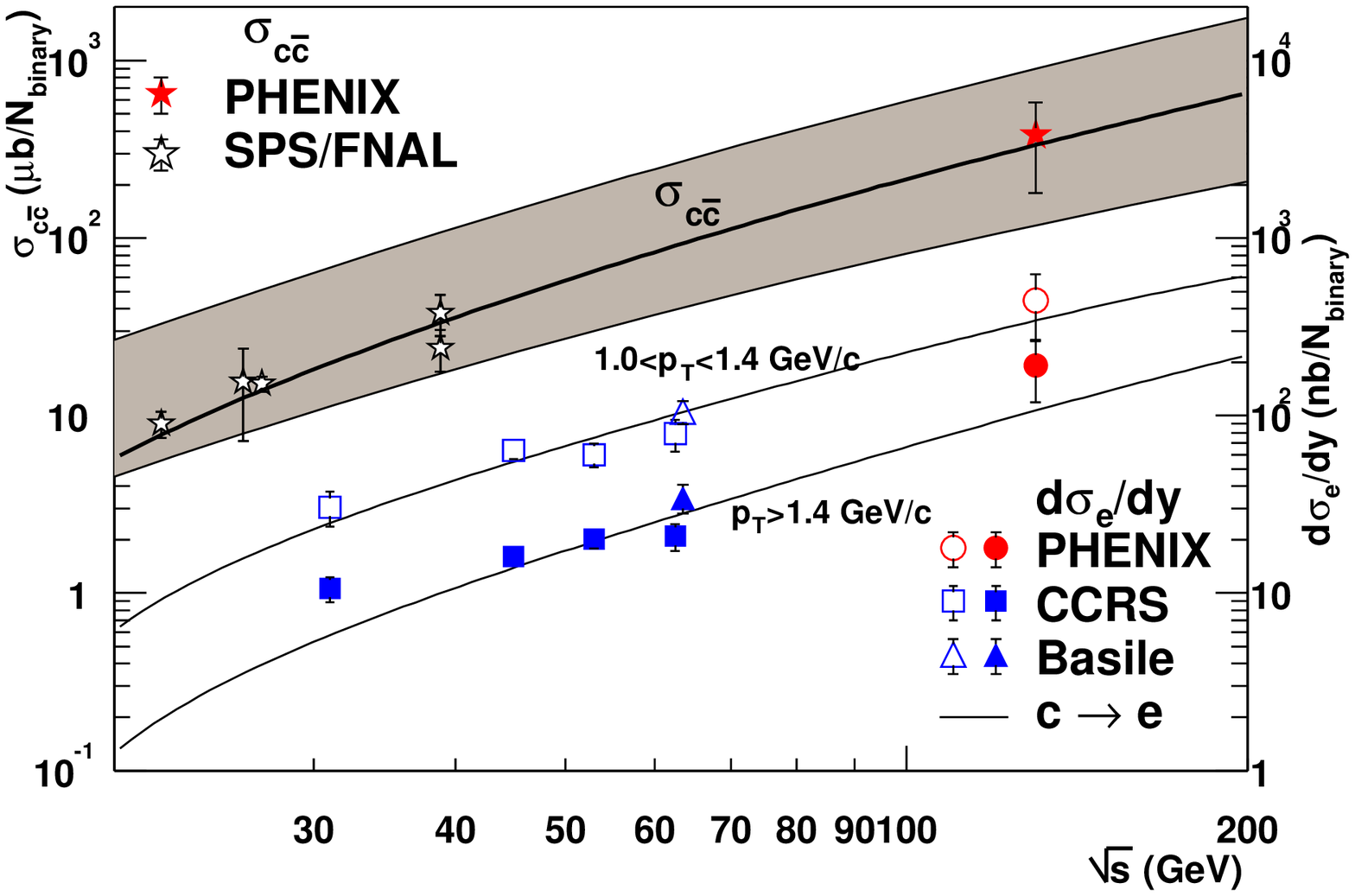, width=75mm}}
 \caption{\it 
PHENIX results on the total charm cross section per binary
collision, compared with lower energy $\pp$ data.
 \vspace{0.5cm}
    \label{fig:phenix-charm-fig4}}
\end{minipage}
\end{figure}

\begin{equation}
\raa = \frac{dN/d\pt({\rm Au+Au})}{\nc \times dN/d\pt({\pbarp})}
\end{equation}

The PHENIX results of $\raa$ vs. $\pt$ for charged hadrons
and neutral pions in 130 GeV Au+Au collisions
are shown in Fig. \ref{fig:phenix-jetq}.
At low-$\pt$, the $\raa$ is $1/6$, as expected for
wounded nucleon scaling (since $\nc/(\np/2) \sim 6$ in central Au+Au).
However, while $\raa$ increases with $\pt$, as might be expected for
the gradual dominance of hard processes, at 2 GeV the rise stops,
with $\raa$ saturating or decreasing, depending on the particle type,
well below unity.
STAR results on $\raa$ show a marked decrease above 2 GeV.
Preliminary results at 200 GeV from both experiments confirm this
decrease and find that above 5-6 GeV, $\raa$ becomes approximately 
constant at $\approx 1/6$ out to 9 GeV (PHENIX all charged and $\pi^o$) and 
$\approx 1/3$-$1/2$ out to 11 GeV (STAR all charged).
These results are in qualitative agreement with the jet quenching
hypothesis and theoretical descriptions are rapidly becoming available.


While large energy loss effects were expected for the nearly-massless
light quarks, heavy quarks are expected to radiate far less.
This is because of the ``dead-cone'' effect, where 
radiation is suppressed at angles less than
$\theta<m_q/E$, where $m_q$ is the mass of the heavy quark, and $E$ is its
energy \cite{dead-cone}.
This expectation has been tested by PHENIX using measurements of
single electrons and positrons in central 130 GeV Au+Au collisions.
They observe a significant excess of electrons
above the expected hadronic and photonic backgrounds and 
attribute it to the presence of open charm,
thus extracting an open charm cross section per binary collision  
$\sigma_{\overline{c}c} = 380 \mu b \pm 200 ({\rm sys}) \pm 60 ({\rm stat})$
\cite{phenix-charm}.
This is consistent with an extrapolation
of FNAL and ISR measurements using a PYTHIA calculation tuned to
reproduce the lower energy results,
as shown in Fig. \ref{fig:phenix-charm-fig4}.

Thus, there seems to be no violation of collision scaling in the 
charm sector, consistent with no energy loss, as expected from
the dead-cone effect.
However, it must be kept in mind that this measurement has
large systematic errors, which are being carefully 
addressed in the higher-statistics 200 GeV data.   
It should also be noted that the electrons have not been
directly tagged as coming from charm decays, e.g. by measuring
the characteristic decay length of the charm mesons.
Future upgrades to the RHIC experiments 
seek to address this issue definitively, which is a crucial
piece of information both for understanding jet quenching and
as a baseline for $J/\Psi$ suppression.

\section{Ultra Peripheral Interactions}

\begin{floatingfigure}[r]{8cm}
\centerline{\epsfig{file=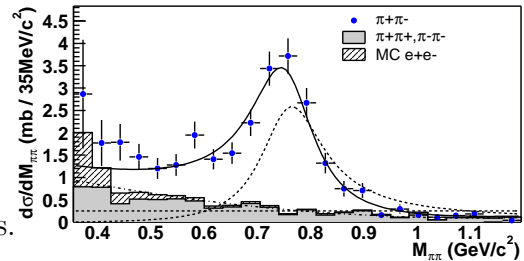, width=7cm}}
 \caption{\it 
STAR results on coherent $\rho^0$ production in
130 GeV ultraperipheral Au+Au collisions.
 \vspace{0.5cm}
    \label{fig:mb_m}}
\end{floatingfigure}

The colliding gold beams at RHIC have also opened up opportunities
of studying the physics of coherent meson production.  The highly
compressed ions act as sources of very strong Coulomb fields as
well as pomerons.  Thus, at large impact parameter,  
one expects photon- photon or photon- pomeron interactions,
as well as virtual photon exchange.  
Vector mesons can be produced coherently by a virtual photon, 
emitted by one nucleus,
fluctuating into a quark-antiquark pair and scattering elastically 
off of the other nucleus, 
provided they satisfy the conditions $\pt < \pi \hbar/R_A$ (90 MeV)
and $p_{||}<\pi\hbar\gamma/R_A$ (6 GeV) where $\gamma$ is the lorentz
boost of the nucleus (70 at RHIC 130 GeV).
The $\rho^o$ production cross section can be predicted
using Glauber extrapolations of $\rho^o$ photoproduction data
to be 350 mb \cite{coherent-rho-theory}.

Using a combination of special triggers and minimum-bias samples,
STAR has extracted a strong coherent $\rho^o$ peak both in 
collisions where the gold nuclei are intact, and where one or
more neutrons is dissociated from each nucleus via coulomb
excitation (shown in Fig. \ref{fig:mb_m}).  They extract
an exclusive $\rho^o$ cross section of $410 \pm 190 \pm 100$ mb
\cite{coherent-rho}.
The agreement between these measurements and the Glauber 
predictions suggests that $\rho^o$ production and Coulomb excitation
are independent processes, and thus factorize.
These measurements thus
point to new opportunities in the physics of strong fields.

\section{Summary of Results}

With the new RHIC data, systematic data now exists for heavy
ion collisions as a function of $\s$ over several orders of
magnitude and as a function of impact parameter.  
These data test the interplay between hard and soft processes
in a large-volume system where nucleons are struck multiple times.
The data is consistent with creating a deconfined state (jet
quenching) that forms quickly (saturation models), expands
rapidly (radial and elliptic flow) and freezes out suddenly
(single freezeout and blast wave fits).  There are also intriguing
connections with particle production in elementary systems, which
point to the role of the energy available for particle production
on the features of the final state.  Many in this field are 
optimistic that the careful understanding of this experimental
data may lead to the theoretical breakthroughs that will connect
these complex systems to the fundamental lattice predictions.

\section{Acknowledgments}

I would like to thank the organizers of Physics in Collision
for inviting me to speak on behalf of the heavy ion community.
I would also like to thank the RHIC experiment 
spokespeople, W. Busza, T. Hallman, 
F. Vidaebeck, and W. Zajc, for their input. 
Finally, special thanks to J. Nagle, C. Roland, and the PHOBOS collaboration 
for valuable suggestions and advice while preparing the talk and manuscript.

\end{document}